\documentstyle[aps,multicol,epsf]{revtex}

\draft
\def\beginwide{
        \end{multicols} \vspace*{-0.5cm} \noindent
        \rule{3.5in}{.1mm}\rule{.1mm}{5mm} \widetext \medskip }
\def\beginwidetop{
        \end{multicols} \vspace*{-0.5cm} \noindent
        \widetext \medskip }
\def\endwide{
        \hspace*{3.35in}~\rule[-5mm]{.1mm}{5mm}\rule{3.5in}{.1mm}
        \begin{multicols}{2} \vspace*{-1.0cm} \noindent }
\def\endwidebottom{
        \begin{multicols}{2} \vspace*{-1.0cm} \noindent }

\newcommand{\beq}{\begin{equation}}
\newcommand{\eeq}{\end{equation}}
\newcommand{\bdis}{\begin{displaymath}}
\newcommand{\edis}{\end{displaymath}}
\newcommand{\bea}{\begin{eqnarray}}
\newcommand{\eea}{\end{eqnarray}}
\newcommand{\barr}{\begin{array}}
\newcommand{\earr}{\end{array}}

\begin{document}

\title{Dynamics of a ferromagnetic domain wall
and the Barkhausen effect}

\author{Pierre Cizeau$^1$, Stefano Zapperi$^1$, Gianfranco Durin$^2$ and
H. Eugene Stanley$^1$}

\address{$^1$Center for Polymer Studies and Department of Physics,
        Boston University, Boston, Massachusetts 02215\\
	$^2$ Istituto Elettrotecnico Nazionale Galileo Ferraris and 
	GNSM-INFM, Corso M. d'Azeglio 42, I-10125 Torino, Italy}

\maketitle

\begin{abstract}
We derive an equation of motion for the
the dynamics of a ferromagnetic domain wall driven
by an external magnetic field through a disordered
medium and  we study the associated depinning
transition.  The long-range dipolar interactions set the
upper critical dimension to be $d_c=3$, so
we suggest that mean-field exponents  describe the
Barkhausen effect for three-dimensional soft ferromagnetic
materials. We analyze the scaling of the Barkhausen jumps
as a function of the field driving rate and the intensity
of the demagnetizing field, and find results
 in quantitative agreement with experiments on
crystalline and amorphous soft ferromagnetic alloys. 
\end{abstract}

\date{\today}

%% 75.60.Ej Magnetization curves, hysteresis, Barkhausen and related effects
%% 75.60.Ch Domain walls and domain structure 
%% 68.35.Ct Interface structure and roughness

\begin{multicols}{2}

The magnetization of a ferromagnet displays discrete jumps
as the external magnetic field is increased.  This
phenomenon, known as the Barkhausen effect,
was first observed in 1919 by recording the tickling noise
produced by the sudden reversal of the Weiss
domains \cite{bark}. The Barkhausen effect has been widely
used as a non-destructive method to test magnetic
materials and the statistical properties of the noise have
been analyzed in detail \cite{oldexp,porte}. In particular,
it has been observed that the distributions of sizes and durations
of Barkhausen jumps 
decay as power laws at low applied field rates
\cite{bdm,durin1,sava}. In addition to  its practical
and technological applications, the Barkhausen effect has
recently attracted a growing interest as an
example of a complex dynamical system displaying critical
behavior \cite{cote,sethna,urbach,weissman,nara}.

In soft ferromagnetic materials the magnetization process is composed of
two distinct mechanisms \cite{herpin}.
 (i) When the field is increased from the saturated
region, domains nucleate in the sample, typically starting from the
boundaries.
 (ii) In the central part of the hysteresis loop, around the
coercive field, the magnetization process is mainly due to domain wall
motion. The disorder present in the material
(due to non magnetic impurities,
lattice dislocations, residual stresses, etc.) is responsible 
for the jerky motion of the domain walls, giving rise to the jumps 
observed in the magnetization.  
The moving walls are usually parallel to the magnetization
(180$^\circ$ walls), and span the sample from end to end \cite{herpin}. 
The statistical properties of the Barkhausen effect are normally
studied in the central part of the hysteresis loop and can therefore be
understood by studying domain wall motion.

It has recently been proposed 
to relate the scaling properties of the Barkhausen noise to the critical
behavior expected at the depinning transition of an elastic
interface \cite{urbach,nara}. 
The numerical values of the scaling exponents, however, do not
agree with most experimental data \cite{bdm,durin1,sava}. 
Interestingly, a quantitative description of the phenomenon
can be obtained by a simple phenomenological model where
the wall is described as a single point moving 
in a correlated random pinning field \cite{abbm}. 

Here, we present
an accurate treatment of magnetic interactions in the context
of the depinning transition, which allows us to explain the experiments
and to give a microscopic justification for the model of 
Ref.~\cite{abbm}. We study a single domain wall
separating two regions with magnetization
of constant magnitude $M$ and opposite directions. 
We assume that the wall surface does not form overhangs, and describe the
position of the wall by its height $h(\vec{r},t)$. The
motion of the wall is overdamped because of eddy currents
so, neglecting thermal fluctuations, 
the evolution of $h(\vec{r},t)$ is governed by 
\beq
\frac{\partial h(\vec{r},t)}{\partial t}=
-\frac{\delta E(\{h(\vec{r},t)\})}{ \delta h(\vec{r},t) },
\label{eq:ovd}
\eeq
where $E(\{h(\vec{r},t)\})$ is the energy of the system \cite{neel1,kron}.
We will show that incorporating  
the effects of ferromagnetic, magnetocrystalline,
magnetostatic and dipolar interactions, and disorder,
the equation of motion is
%\beginwide
\bea
\nonumber
\lefteqn{\frac{\partial h(\vec{r},t)}{\partial t} =
\nu_0 \nabla^2h(\vec{r},t)+H-H_d} \\
%\eeq
%\beq
& &+\int d^2r^\prime K(\vec{r}-\vec{r}^{\;\prime})
(h(\vec{r}^{\;\prime},t)-h(\vec{r},t))+\eta(\vec{r},h).
\label{eq:tot}
\eea
%\endwide
Here $\nu_0$ is the surface tension, $H$ is the external magnetic field,
the demagnetizing field $H_d$ and  the non-local term are 
due to dipolar interactions, 
and we model the disorder with a random force, with Gaussian distribution 
and short range correlations
\beq
\langle\eta(\vec{r},h)\eta(\vec{r}^{\;\prime},h^\prime)\rangle
=\delta^2(\vec{r}-\vec{r}^{\;\prime})\Delta(h-h'),
\eeq
where $\Delta(x)$ decays very rapidly for large values of the
argument.

The dipolar interactions are treated considering
effective magnetic charges induced by the discontinuities of
the magnetization across the boundaries
of the sample and the domain
wall \cite{jackson}. 
The corresponding magnetic surface charge is given by
\beq
\sigma = (\vec{M}_1-\vec{M}_2)\cdot\hat{n},
\eeq
where $\hat{n}$ is normal to the surface and $\vec{M}_1$ and $\vec{M}_2$ are
the magnetization vectors on each side of the surface. 
The surface charge induced at the boundary of the sample gives rise to
a demagnetizing field which opposes the external field \cite{herpin}.
The simplest approximation is to consider this field to be constant
throughout the sample, and to be proportional to the total magnetization
\cite{urbach,abbm}. This yields the term $H_d=-k\int d^2r~h(\vec{r},t)$
in Eq.~(\ref{eq:tot}),
where the demagnetizing factor $k$ takes into account
the geometry of the domain structure, the shape of the system
and its size. 

\begin{figure}[htbp]
\narrowtext
\centerline{
        \epsfxsize=7.0cm
        \epsfbox{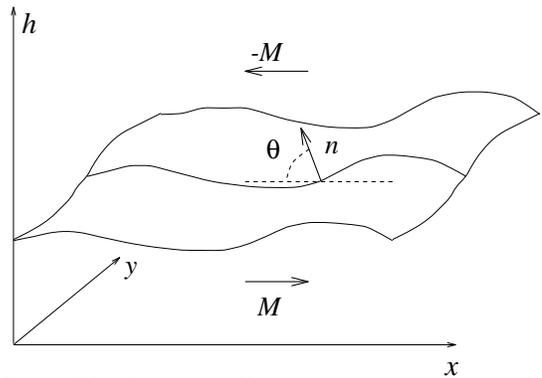}
%        \vspace*{0.5cm}
        }
\caption{ The domain wall separating
two regions of opposite magnetization.
The discontinuities of the normal component of the magnetization
across the domain wall generate magnetic charges.}
\label{fig:1}
\end{figure}

Magnetic surface charges will also appear on the domain wall
when its surface is not parallel to the magnetization.
In the limit of infinite anisotropy \cite{neel_footnote} and small bending of
the surface, we can express the surface charge as
\beq
\sigma(\vec{r}) = 2M \cos\theta \simeq 
2M\frac{\partial h(\vec{r},t)}{\partial x},
\eeq
where $\theta$ is the local angle between the vector normal
to the surface and the magnetization (see Fig.~\ref{fig:1}). 
The interaction energy of these charges $E_d =(1/2)
\int d^2r d^2r^{\prime} \sigma(\vec{r})\sigma(\vec{r}^{\;\prime})
/|\vec{r}-\vec{r}^{\;\prime}|$
gives rise to the non-local kernel present in Eq.~(\ref{eq:tot})
\cite{neel1,kron,long}:
\beq
K(\vec{r}-\vec{r}^{\;\prime})=
\frac{2M^2}{|\vec{r}-\vec{r}^{\;\prime}|^3}\left(1+
\frac{3(x-x^\prime)^2}{|\vec{r}-\vec{r}^{\;\prime}|^2}\right).
\label{eq:ker}
\eeq
The interaction (\ref{eq:ker}) is long-range  and anisotropic, as can
also be seen by considering the Fourier transform
\beq
K(p,q) =\frac{2M^2}{\pi} \frac{p^2}{\sqrt{p^2+q^2}},
\eeq 
where $p$ and $q$ are the two components of the Fourier
vector. Moreover, an estimate of the order of magnitude
of $K(\vec{r}-\vec{r}^{\;\prime})$ shows that it dominates over the
surface tension for all
 length scales of interest \cite{ordre}.

Apart from the non-local kernel, Eq.~(\ref{eq:tot}) is the
equation proposed in Ref.~\cite{urbach} which, in
turn, reduces when $k=0$ to a driven elastic interface in
the presence of quenched disorder \cite{koplev,natt,nf}. 

For $k=0$ (no demagnetizing field), Eq.~(\ref{eq:tot})
displays a depinning transition: i.e. there exists a critical field
$H_c$ such that for $H<H_c$ the interface is pinned while for $H>H_c$,
it moves with non-zero average velocity.
At $H=H_c$, the system exhibits scaling
properties: the interface moves
by avalanches whose sizes $s$ and durations $T$ distributions follow
 power laws
\beq
P(s)\sim s^{-\tau},~~~~P(T) \sim T ^{-\alpha}.
\eeq
When $k>0$, the demagnetizing field 
provides an additional restoring force that keeps the interface at
the depinning transition \cite{urbach,nara} if the external field
is increased adiabatically.

Using the functional renormalization group scheme
introduced in Ref.~\cite{cdw}, we find
that---due to the long-range kernel $K(\vec{r}-\vec{r}^{\;\prime})$ in 
Eq.(\ref{eq:ker})---the critical behavior 
of Eq.~(\ref{eq:tot}) differs from
that of elastic interfaces: the
upper critical dimension becomes $d_c=3$ 
\cite{long,ertas,nara_note}, 
instead of $d_c=5$ \cite{natt,nf}. 
Hence we predict that for $d=3$, the motion of the domain wall will be 
described by mean-field theory  (apart from logarithmic
corrections), which yields \cite{long,nf}  $\tau=3/2$,
$\alpha=2$, and that the surface is  flat (the roughness exponent $\zeta$
is zero).  These values differ significantly
from the results of elastic interfaces which in $d=3$ are
$\tau\simeq 1.3$, $\alpha \simeq 1.5$ and $\zeta \simeq 0.7$ 
\cite{nf,heiko}.

Next we make contact between our
approach and the conventional approach that reduces the domain
wall to a single point moving in a random pinning field
\cite{porte,abbm,neel2}. To this end, we introduce an
infinite range  version  ($ d\to\infty$) of Eq.~(\ref{eq:tot}), which
should have the same critical behavior as Eq.~(\ref{eq:tot})
but has the advantage of being much simpler to analyze.

To treat the infinite-range model, we discretize the
interface and consider that all $N$ elements are at the same distance
from
each other. Eq.~(\ref{eq:tot})
then becomes \cite{fisher}
\beq
\frac{\partial h_i(t)}{\partial t}=
H(t)-\chi\bar{h}+J(\bar{h}-h_i(t))
+\eta_i(h),
\label{eq:mf2}
\eeq
where $\bar{h}\equiv\sum_{i=1}^{N} h_i/N$, $\chi\equiv Nk$, 
$J\equiv (\nu_0+2M^2)$, and the external
field $H(t)$ increases at a finite constant rate.  Summing
Eq.~(\ref{eq:mf2}) over all sites $i$, we obtain an equation
for the total magnetization $m\equiv N\bar{h}$
\beq
\frac{dm}{dt}=\tilde c t-\chi m +\sum_{i=1}^{N} \eta_i(h),
\label{eq:abbm}
\eeq
where the time dependence of the field has been made
explicit. We can approximate $\sum \eta_i$ by an effective
random pinning field $W(m)$, depending only on the magnetization.
When the interface moves between two 
configurations, the change in $W$ is 
\beq W(m^{\prime})-W(m) =
{\sum_i}^{\;\prime} \Delta \eta_i, 
\eeq 
where the sum is restricted 
to the sites that have moved (i.e. their
disorder is changed).  The total number of such sites
scales, on average, as $|m^{\prime}-m|$, since for $d \geq d_c$
the size of an avalanche scales like its area
\cite{long,nf}. Assuming that the $\Delta\eta_i$ are
uncorrelated and have random signs, we find that the effective pinning field 
is correlated,
\beq
\langle |W(m^{\prime})-W(m)|^2 \rangle = D |m^{\prime}-m|,
\label{eq:W}
\eeq
where $D$ sets the scale of  the fluctuations of $W$.
These correlations of ``Brownian'' type have been experimentally observed in
SiFe alloys \cite{porte}. In the model of  Alessandro {\it et al} \cite{abbm}
(ABBM), the domain wall is treated as a single point moving in a Brownian
correlated field. The ABBM model is equivalent to
Eqs.(\ref{eq:abbm}) and (\ref{eq:W}), and predicts that the avalanche
exponents should depend on the field driving rate, with
$\tau=3/2-c /2$ and $\alpha=2-c$ \cite{durin1}, where $c\equiv \tilde c /D$.  
In the adiabatic limit
$c\to 0$, we recover the mean-field exponents of the depinning
transition.  Moreover, our results imply that Brownian correlations in
the pinning field do not reflect peculiar long-range correlations in the
impurities, but represent an {\em effective} description of the
collective motion of the interface.

To confirm the results obtained above, we first  simulate an automaton
version of Eq.~(\ref{eq:tot}) in three dimensions, applying 
an adiabatically increasing magnetic field. 
In this limit, we recover, for the distribution of avalanche sizes
and durations, the 
results expected for the depinning transition
in mean-field theory \cite{long}.

\begin{figure}[htbp]
\narrowtext
\centerline{
        \epsfxsize=7.0cm
        \epsfbox{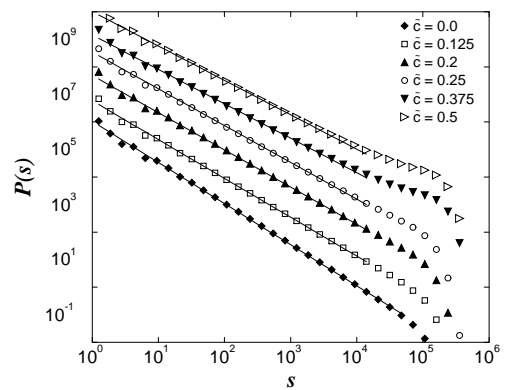}
%        \vspace*{0.5cm}
        }
\caption{The distribution of avalanche sizes in the infinite range
model as a function of $c$ for $N=32696$, $\chi=0.0075$.
The lines are the theoretical predictions $\tau=3/2-c/2$. }
\label{fig:3}
\end{figure}

\begin{figure}[htbp]
\narrowtext
\centerline{
        \epsfxsize=7.0cm
        \epsfbox{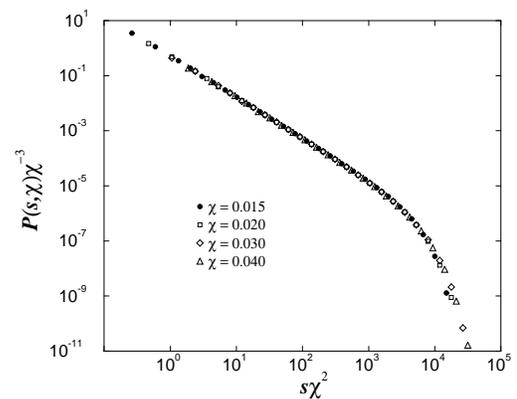}
%        \vspace*{0.5cm}
        }
\caption{The distribution of avalanche sizes in the infinite range
model as a function of $\chi$ for $N=32696$ and $\tilde c =0$. 
The data collapse is
obtained from the scaling function $P(s,\chi)=s^{-3/2}f(\chi^2 s)$.}
\label{fig:new}
\end{figure}

To overcome the numerical limitations posed by
Eq.~(\ref{eq:tot}) and
obtain an extensive avalanche statistics as a function
of $c$ and $\chi$, 
we next simulate the infinite-range model of Eq.(\ref{eq:mf2}). 
We measure the distributions
of domain wall velocities and avalanche sizes and
durations (see Figs.~\ref{fig:3} and \ref{fig:new}, and \cite{long}). 
For $c<1$, we find power laws with exponents in
perfect agreement with the ABBM model.
The cut-off of the distributions is
determined by the demagnetizing fields and diverges for $\chi \to 0$
 \cite{long}. In the case of the avalanche
size distribution, it scales as  $\chi^{-2}$ (see Fig.~\ref{fig:new}).
It would be interesting to experimentally study the
behavior of the cut-off by externally 
changing the demagnetizing field.

Our results are in agreement with 
experiments on crystalline (SiFe) \cite{bdm,durin1} and
amorphous (Co-base and Fe-base) ferromagnetic alloys
\cite{durin3} performed at different values of $c$, 
which yield $\tau=3/2-c/2$ and $\alpha=2-c$.
A direct comparison with experiments in which 
the parameter $c$ is not defined, as in the case of a 
sinusoidal driving field, is problematic \cite{sava,cote}.
The result presented in Ref.~\cite{urbach} (i.e. $\tau \simeq 1.3$)
could be due to a finite driving rate ($c\simeq 0.4$). 
In addition, it is important to remark that the present theory applies only
if domain wall motion is the dominant magnetization process.
This was indeed the case in Refs.~\cite{bdm,durin1,durin3},
where the noise was recorded only in the region of constant
permeability around the coercive field. 
A detailed critical discussion of the experimental results
reported in the literature can be found in Ref.~\cite{durin3}.

In particular geometries, typically frames or toroid samples, the
demagnetizing field is absent ($\chi=0$) \cite{porte}. It
is then possible to observe experimentally the depinning
transition of domain walls. It has been reported 
that the average velocity of the domain walls, 
in different ferromagnetic materials \cite{beta=1}, 
increases for $H>H_c$, as
\beq 
v \sim
(H-H_c)^\beta,
\eeq 
with $\beta=1$, in agreement with the theory 
($\beta=1$ is expected in mean-field
theory \cite{nf}).

We have seen that avalanche distributions can be described by power laws
with exponents that do not depend on material details.  The power
spectrum of the noise displays instead a more complex structure and does
not show such a marked universality. At low frequency the power spectrum
grows with an exponent varying between $\psi=0.5$ for crystalline
alloys and $\psi=1$ for amorphous alloys, while at
high frequencies it decays with an exponent varying between $\psi=-2$
for crystals and $\psi=-1.6$ for amorphous alloys
\cite{oldexp,bdm,durin1,sava,durin3}.  
In samples with a {\em single} domain
wall present, the power spectrum was found to
decay as $\omega^{-2}$ \cite{porte}.

Following the analysis of Tang et al. \cite{tang},
we obtain at low frequency $\psi=1$ for the power spectrum measured on a
single site and $\psi=0$ when the signal is averaged over the whole
system. For the averaged spectrum, we also find a $\omega^{-2}$ decay
at large frequencies, due to the Brownian properties of the effective
pinning field. The discrepancies between theory and
experiments could be due to the presence of many domain walls
interacting through the demagnetizing field. When a domain wall
starts to move, the demagnetizing 
field increases, creating a larger pinning
force on the other walls. Therefore, on short time scales 
the interactions between the walls is irrelevant and should not 
change the avalanche distributions.
On larger time scales, this effect may be important
and could modify the properties of the power spectrum.
In order to clarify this issue, it would be necessary
to analyze in detail the dynamics of many coupled domain
walls. 

The present theory for the Barkhausen effect, based on the
depinning of a ferromagnetic domain wall, should apply to soft
ferromagnetic materials, which are frequently used in experimental
studies of the Barkhausen
effect \cite{oldexp,porte,bdm,durin1,sava,urbach,abbm}.   For 
hard ferromagnets and rare earth materials where
strong random anisotropies prevent the formation of domains, 
disordered spin models could be appropriate \cite{sethna}.

We thank G. Bertotti, J.-P. Bouchaud, D. Ertas, M. M\'ezard,
S. Milo\v sevi\'{c} and S. Roux
for useful discussions, and L. A. N.  Amaral and M. Meyer 
for critical reading of
the manuscript. The Center for Polymer Studies is supported by NSF.

%\pagebreak

%\pagebreak

\end{multicols}

\begin{thebibliography}{99}
\bibitem{neel_footnote}
	In the case of finite anisotropy, the direction of
	the magnetization can deviate from the easy axis. This
	effect can be included in the derivation, and the resulting
	kernel is qualitatively unchanged \cite{neel1}.
\bibitem{ordre}
	The long-range kernel become relevant for length
	scales larger than $L\sim \nu_0/M^2$. In typical
	ferromagnets, $M \sim 10^3$ and $\nu_0 \sim 1$ (in cgs
	units) (see page~713 of \cite{herpin}). 
	This implies $L~\sim 10^{-6} \mbox{ cm}$, 
	which is of the order of the
	domain wall thickness.
\bibitem{nara_note}	
	Narayan \cite{nara} has considered dipolar interactions,
	concluding that they do not change
	the universality class of elastic interfaces,
	in contrast with our results.
\end{thebibliography}
\end{document}